\documentclass[
    ,final            % use final for the camera ready runs
  ]{aipproc}

\usepackage{psfrag,euscript,amsmath}

\layoutstyle{6x9}

\begin{document}

\title{\boldmath $B$-decays in the heavy-quark expansion\unboldmath\footnote{Talks presented at the 15th Topical
    Conference on Hadron Collider Physics, HCP2004, Michigan State
    University, June 14-18, 2004 and at the 11th International QCD
    Conference, Montpellier, July 5-10, 2004.}}

\author{Thomas Becher}{
  address={Fermi National Accelerator Laboratory,\\
P.~O.~Box 500, Batavia, IL 60510, USA}
}

\begin{abstract}
Progress in the theoretical description of $B$-meson
decays, in particular decays to light hadrons, is reviewed. The factorization
properties of such decays can be analyzed using the soft-collinear
effective theory. Applications of the effective theory to
both inclusive and exclusive decays are discussed.
\end{abstract}

\maketitle

\begin{flushright}
\raisebox{9cm}[0ex][0ex]{
{\small
\hfill\parbox{5cm}{
{\textsf{FERMILAB-CONF-04-309-T}}\\
{\textsf{hep-ph/0411065}}}}
}\vspace*{-5ex}
\end{flushright}

%%%%%%%%%%%%%%%%%%%%%%%%%%%%%%%%%%%%%%%%%%%%
%% MAINMATTER
%%%%%%%%%%%%%%%%%%%%%%%%%%%%%%%%%%%%%%%%%%%%

%\renewcommand{\thefootnote}{\arabic{footnote}}
%\setcounter{footnote}{0}

\section{Introduction} 

The first flavor physics conference I attended was devoted to studying
the possibility of a future, second generation, high-luminosity
$B$-factory. The title of the conference held in the year 2000 was
``Beyond $10^{34}$ $e^+e^-$ workshop''. Four years later the current
$B$-factories at SLAC and KEK have reached luminosities of
$10^{34}cm^{-2}s^{-1}$ (the record luminosity, $1.4 \times
10^{34}cm^{-2}s^{-1}$, is held by the KEKB accelerator). The
impressive performance of the experiments has made it possible to
apply methods that were thought to be reserved for future machines,
e.g.~it is now possible to obtain constraints on the CKM angle
$\alpha$ from decays $B\rightarrow \pi\pi$ and $B\rightarrow \rho\rho$
using only the approximate isospin symmetry of the strong interaction
as an input.  Methods which use data to eliminate all strong
interaction physics, like the isospin analysis of the $B\rightarrow
\pi\pi$ decays \cite{Gronau:1990ka}, have the advantage of being
conceptually simple and therefore well controlled.  The disadvantage
of such techniques is that they are experimentally extremely demanding
and for this reason the bounds obtained are not (yet) very stringent.
More importantly, such methods can only be used to establish new
physics but not to explore it, since they rely on the specific
structure of the weak interaction in the Standard Model (SM).

In this talk I focus on methods which are more ambitious in that they
evaluate part of the strong interaction effects in $B$-decays. All of
these tools rely on an expansion of the decay amplitudes in inverse
powers of the $b$-quark mass. The predictions obtained are then
accurate up to terms suppressed by powers of the heavy-quark mass.
The classic example is heavy-quark effective theory (HQET), which has
allowed a determination of $|V_{ cb}|$ at the level of a few per cent
from exclusive semi-leptonic decays. However, HQET is not applicable
in $B$-decays in which some of the outgoing, light particles have momenta of
the order of the $b$-quark mass, since momenta of this magnitude have
been integrated out in the construction of the effective theory.
Because of this restriction HQET can in general not be used to analyze
$B$-decays involving light hadrons, such as $B\rightarrow \pi\pi$,
$B\rightarrow K^*\gamma$ or $B\rightarrow
\pi\ell\nu$.  However, since they probe small CKM elements and flavor changing
neutral currents such decays are of great interest to test the
Standard Model.

While HQET is not applicable, it is still true that these processes
simplify in the heavy-quark limit: the larger the heavy-quark mass,
the larger the energies of the outgoing light mesons become and the
methods that are used to establish QCD factorization theorems at large
momentum transfer become applicable. In this case the hard scale is
set by the mass of the heavy $b$-quark, and factorization theorems for
inclusive \cite{Neubert:1993ch, Bigi:1993ex, Korchemsky:1994jb} as
well as exclusive \cite{Beneke:1999br} $B$-decays to light hadrons
arise in the heavy-quark limit.  In the past few years, an effective
field theory framework for the analysis of such decays has been
developed
\cite{Bauer:2000yr,Bauer:2001yt,Chay:2002vy,Beneke:2002ph,Hill:2002vw,Becher:2003qh}:
the soft-collinear effective theory (SCET) gives the heavy-quark
expansion of heavy-to-light decays and permits the study of their
factorization and renormalization properties.

SCET turns out to be more involved than HQET. One complication is that
the low energy theory involves multiple scales.  In order to get a
simple counting of powers of the expansion parameter in such a
situation, a single QCD quark (or gluon) field is represented by
several effective theory fields whose momentum components scale
differently with the heavy-quark mass $m_b$. The different effective
theory fields are also called modes and correspond to
momentum regions, where the QCD diagrams for a given process develop
singularities in the heavy-quark limit. In addition to the soft fields
present in HQET, SCET contains collinear fields which have large
energy and carry large momentum in the direction of the outgoing light
hadrons. Another complication is that the non-perturbative input needed
in  applications of SCET is functional. This is quite generally the
case for factorization in hard processes; the most familiar example of
such functional input are the parton distribution functions that are
needed for the calculation of hadron-hadron scattering processes.  The
non-perturbative functions relevant in the case of $B$-decays are the
light-cone distribution amplitudes of the $B$-meson and the light mesons
for exclusive decays and the shape function for inclusive
decays.

In my talk I first briefly review HQET and present it in the same
diagrammatic language that is used in the construction of SCET. I then
discuss the effective theory for inclusive decays to light hadrons,
called SCET$_{\rm I}$, and sketch how the factorization theorem for
these decays is obtained. The effective theory for exclusive decays,
SCET$_{\rm II}$, is based on the same basic formalism, but contains
different degrees of freedom. In applications, one often performs two
matching steps. First one matches QCD onto SCET$_{\rm I}$ at a scale
of the order of the $b$-quark mass.  At a lower scale $\mu^2\sim m_b
\Lambda$ between the heavy-quark mass and a typical hadronic scale
$\Lambda$, one matches SCET$_{\rm I}$ onto SCET$_{\rm II}$. As an
example of an exclusive decay which has been analyzed using this two-step procedure, I discuss the factorization properties of the
heavy-to-light form factors.

\section{The heavy-quark expansion}

Weak decays involve a large number of very disparate scales: from very
large scales like the mass of the $Z$- and $W$-bosons, and the top
quark mass down to very small ones such as the light-quark or the
neutrino masses. There are strong interaction effects associated with
all of these scales and the effects associated with large scales can
be calculated in perturbation theory. To separate the calculable
high-energy QCD effects from the non-perturbative low energy physics,
one performs expansions in small ratios of the scales in the
problem. These expansions are most easily performed by using
effective theories which are obtained by integrating out the physics
associated with the large scales.

The first step is to expand the weak interactions in $\Lambda/M_W\sim
\Lambda/M_Z \sim \Lambda/m_t $, where $\Lambda$ is a momentum
component or the
mass of one of the lighter particles participating in the weak decay. As a
result of this expansion, one obtains the low-energy effective weak
Hamiltonian which has the form
\begin{equation}
{\cal H}=\frac{G_F}{\sqrt{2}}\sum_i V^i_{\rm CKM} C_i(\mu) Q_i \,.
\end{equation}
The operators $Q_i$ are local operators built from the light
fields only: since the energies involved in weak decays of hadrons are
much smaller than the $W$-, $Z$- and top-mass, these particles are
always highly virtual and do not appear as dynamical degrees of
freedom at low energies. The operators $Q_i$ are multiplied by
the CKM factors $V^i_{\rm CKM}$ and Wilson coefficients $C_i(\mu)$. The Wilson
coefficients encode the QCD effects from higher energies. They have
been calculated accurately at next-to-leading order in renormalization
group improved perturbation theory \cite{Buchalla:1995vs}.  The
effective weak Hamiltonians can be viewed as a generalization of the
Fermi theory to include all quarks and leptons and their electroweak
and strong interactions. The rich phenomenology of hadronic and
radiative weak decays (as well as some of the difficulties in their
theoretical analysis) result from the fact that the effective
Hamiltonian for such decays consists of many operators and a single
process can thus involve several different weak decay mechanisms.

For kaon decays the remaining low-energy physics is non-perturbative
and one cannot proceed further within perturbation theory. The
situation is different for decays of $B$-mesons: the strong coupling
constant at the mass of the $b$-quark is $\alpha_s(m_b)\approx 0.2$ so
that part of the decay is calculable in perturbation theory. To
separate the calculable part one performs an expansion in $1/m_b$, the
heavy-quark expansion.

\subsection{Heavy-to-heavy decays: HQET and the determination of
  {\boldmath $|V_{cb}|$\unboldmath}}

The heavy-quark expansion for $B$- to $D$-meson decays was constructed
quite some time ago \cite{Neubert:1993mb}. In the meantime these
methods have led to determinations of $|V_{cb}|$ at an accuracy of a
few per-cent, from inclusive as well as exclusive measurements.
The exclusive determination from the semi-leptonic decay
$B^0\rightarrow D^{*}\,\ell \nu$ gives \cite{HFAG}
\begin{equation}
|V_{ cb}|=(41.5\pm 1.0_{\rm exp} \pm 1.8_{\rm theo})\times 10^{-3}\,.
\end{equation}
To reach an accuracy of 4\% in the calculation of an exclusive
hadronic decay is quite an achievement. There are several reasons for
this success. First of all, it turns out that in the heavy-quark
limit, the relevant $B\to D^*$ form factor is equal to unity at zero
recoil. This prediction receives radiative corrections, suppressed by
$\alpha_s(m_Q)$ and non-perturbative power corrections suppressed by
$\Lambda/m_Q$, where $\Lambda$ is a typical hadronic scale and
$m_Q$, with $Q=c,b$, is the heavy-quark mass. Both of these corrections
are known. The perturbative corrections were calculated to
two loops \cite{Czarnecki:1997cf}. By virtue of Luke's theorem
\cite{Luke:1990eg}, the first order non-perturbative corrections
vanish and the terms of order $\Lambda^2/m_Q^2$ were evaluated with a
quenched lattice simulation \cite{Hashimoto:2001nb}.  Preliminary
unquenched results are now also available \cite{Okamoto:2004xg}.

The accuracy is even higher for the inclusive $b\to c$ decay. A
recent global fit to all available data yields \cite{Bauer:2004ve}
\begin{equation}
|V_{ cb}|=(41.9\pm 0.6 )\times 10^{-3} \,.
\end{equation}
The explanation for the tiny uncertainty is similar to the exclusive
case: in the heavy-quark limit the decay rate is given by the
$b$-quark decay rate and the leading corrections of order $\alpha_s$
or $\Lambda^2/m_b^2$ to this predictions have also been evaluated. The
non-perturbative corrections to the decay rate are given by the
$B$-meson matrix elements of the two operators that appear in the
sub-leading HQET Lagrangian. One of these matrix elements can be
related to the the experimental value of the mass difference between
pseudoscalar and vector $B$-mesons.  The second one as well as the
$b$- and $c$-quark masses can be obtained from the experimental
results for moments of the decay spectrum, which depend on the same
parameters as the total rate. The analysis of these moments has been
done independently by two groups \cite{Bauer:2002sh,Gambino:2004qm}
who at this point accuse each other in turn of underestimating their
theoretical uncertainties and algebraic mistakes
\cite{Bauer:2004ve,Uraltsev:2004ra}.

Before discussing the heavy-to-light hadron decays, I would like to
show how the HQET Lagrangian can be derived by expanding QCD Feynman
diagrams around the heavy-quark limit. The same diagrammatic
techniques are used in the construction of the SCET Lagrangian,
discussed in the next section.  At tree level the expansion
is achieved after realizing that at low energies, the heavy quark is
close to its mass shell so that the heavy-quark momentum can be
written as $p_Q=m_Q v_\mu+r_\mu$, where $v^2=1$ and all components of
the residual momentum $r_\mu$ are small compared to $m_Q$. The
expansion of the heavy-quark propagator then reads
\begin{equation}\label{eq:expansion}
\frac{p_Q\!\!\!\!\!\!\!\!\!/\,\,\,+r\!\!\!\!/+m_Q}{(p_Q+r)^2-m_Q^2}=\frac{1+v\!\!\!\!/}{2}
\frac{1}{v \cdot r} + \dots \,.
\end{equation}
Note that ${\rm P}_+=\frac{1+ v\!\!\!\!/}{2}$ is a projector, ${\rm
  P}_+^2={\rm P}_+$.  Since
it appears sandwiched between heavy-quark propagators, the quark-gluon
vertex also simplifies and becomes
\begin{equation}
{\rm P}_+ \,ig\,A\!\!\!\!\!/\, {\rm P}_+ = ig\,v\cdot A\, {\rm P}_+\,.
\end{equation}
It is easy to write a Lagrangian that yields the above propagator and
vertex, namely
\begin{equation}\label{eq:HQET}
{\cal L}_{\rm HQET}=\bar h\, {\rm P}_+\,i\, v\cdot D\, h\,.
\end{equation}
The field $h$ is the heavy-quark field in HQET. Since the Lagrangian
involves the projector ${\rm P}_+$, the field $h$ is is effectively
only a two component field.  

For loop graphs, simply expanding the
heavy-quark propagators in the integrand does not give the correct
result for the expansion of the integral. The problem is that the loop
momentum flowing through the heavy-quark propagator is large in some
momentum regions, so that the expansion (\ref{eq:expansion}) is not
always appropriate.  However, in dimensional regularization the
prescription to obtain the correct result is remarkably simple. What
one needs to do is to perform two different expansions: first, one
assumes the loop momentum to be small, of order $r_\mu$, expands the
integrand and then performs the loop integration without any
restriction on the integration. The contribution obtained in this way is
called the soft part of the integral.  In general, the soft part has
ultraviolet divergences not present in the original integral. These
extra divergences are regulated by dimensional regularization and
manifest themselves as poles in $d-4$.  One then expands the loop integrand a
second time, this time assuming all components of the loop momentum to
be large, of the order of the $b$-quark mass. This second part is
called the hard part of the integral and it has infrared divergences
that are again regulated dimensionally. In the sum the
infrared divergences of the hard part cancel against the ultraviolet
divergences of the soft part and expansion of full integral is
recovered after adding the hard and the soft part. This technique for
performing asymptotic expansions of integrals is also called the
strategy of regions \cite{Smirnov:2002pj}.

\begin{figure}
\includegraphics[height=.23\textheight]{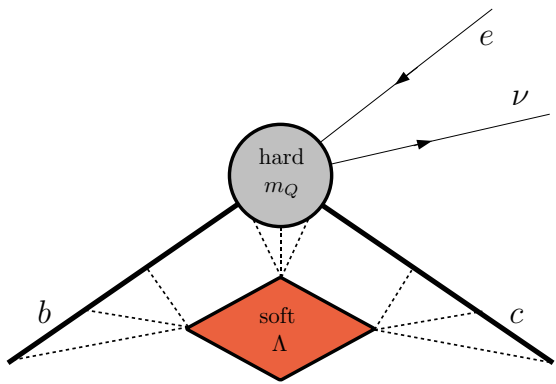}
\caption{Semi-leptonic $B\to D$ decay.  The thick lines denote the
  heavy quarks. The dotted lines represent soft light quarks and gluons.\label{fig:HQET}}
\end{figure}

While the contribution of the soft region corresponds to the effective
theory Feynman diagrams, the hard part is absorbed into the Wilson
coefficients of the effective theory operators.  For the Lagrangian
($\ref{eq:HQET}$) of a single heavy quark at leading order, the only
effect of the hard contributions is to change the normalization of the field
$h$.\footnote{This is only true if $m_Q$ is chosen to be the heavy
  quark pole mass. For any other choice, the hard contribution
  generates a residual mass term of the form $\delta m\, \bar h\,{\rm
    P}_+\,h$ that needs to be included in ${\cal L}_{\rm HQET}$.}
Since the normalization is not a physical observable, it is not really
necessary to introduce a Wilson coefficient to reproduce this effect.

To analyze the exclusive semileptonic $B\rightarrow D$ decay, one
introduces two copies of the heavy-quark Lagrangian (\ref{eq:HQET}),
one for the $b$-quark and another one for the $c$-quark. The
four-velocity vector $v$ needed to remove the large part of the
heavy-quark momenta will be different for the two quarks.  However, 
also in this case only two momentum regions, hard and soft, are
relevant, as illustrated in Figure \ref{fig:HQET}. The
hard-corrections are given by the Wilson coefficient of the
heavy-to-heavy current operator and depend on the scalar product of
the two heavy-quark velocities \cite{Falk:1990cz,Neubert:1991tg} .

\subsection{Inclusive heavy-to-light decays: SCET$_{\rm I}$}

In order to obtain the decay rate of the inclusive semi-leptonic
decay, one uses the optical theorem and calculates the imaginary part
of the forward matrix element of two weak currents
\begin{equation}\label{eq:Tmunu}
T^{\mu\nu}=i\,\int d^4 x\, e^{i q x}\, \langle \bar B(p_B)| \mbox{\bf
  T} \left\{ J^{\dagger\mu}(0) , J^\nu(x) \right\} | \bar B(p_B) \rangle\, ,
\end{equation}
where $J^\mu= \bar u \gamma^\mu\,(1-\gamma_5) b$. To calculate the
total rate, one then uses the Operator Product Expansion (OPE) which
corresponds to the expansion of the total rate around the heavy-quark
limit \cite{Chay:1990da}.  At leading order in the expansion one finds
that the total rate of the meson decay is equal to the free heavy
quark decay rate up to calculable radiative corrections.

\begin{figure}
  \includegraphics[height=.26\textheight]{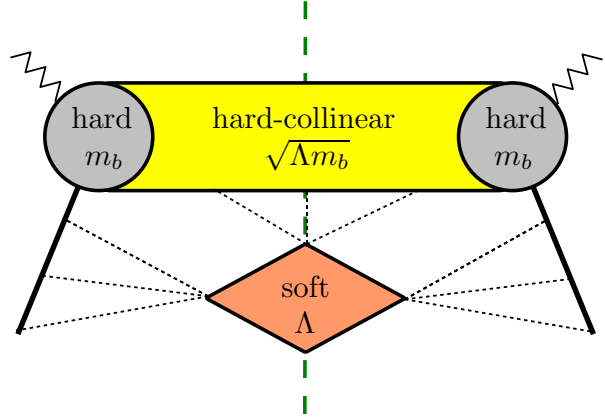}\\
\caption{Inclusive semi-leptonic $B$-decay. The differential decay rate is obtained from the imaginary part of the
  forward matrix element of two weak currents. The thick line denotes
  the heavy $b$-quark. The dotted lines represent soft quarks and
  gluons.\label{fig:scetI}}
\end{figure}

Unfortunately, it is experimentally necessary to impose rather severe
kinematic cuts to discriminate the $B\rightarrow X_u \ell\nu$ and
$B\rightarrow X_s \gamma$ decays against the background. Such cuts
typically enforce a small invariant mass of the system of outgoing
hadrons $M_X^2\sim \Lambda\, m_b\ll E_X^2\sim m_b^2$ and lead to a
breakdown of the OPE. In the heavy-quark limit, the enhanced higher
order terms can be resummed into a non-perturbative shape function $S$
and for this reason the region of small $M_X$ is also called the
shape-function region \cite{Neubert:1993ch,Bigi:1993ex}.  The same
function appears in the semi-leptonic and the radiative decays and one
obtains a relation between the decay rates. To perform the perturbative
analysis of the factorization properties of the partial decay rate in
the shape function region, one replaces the in- and outgoing mesons
with interpolating currents and studies the vacuum expectation value
of the correlator of the weak current with the two meson currents. The
structure of the diagrams contributing to this correlator is shown in
Figure \ref{fig:scetI}.

It turns out that the hard and soft momentum regions are not
sufficient to obtain the heavy-quark expansion of the diagrams
relevant for the heavy-to-light decays.  The reason is that the decay
of the heavy quark typically produces a jet of energetic light
particles. Since their energies are of the order of the $b$-quark
mass they cannot be described by the soft fields present in HQET and
require the introduction of additional, so-called collinear fields. In
HQET one uses the reference unit vector $v_\mu$ to isolate the large
part of the $b$-quark four-momentum, a common choice is $v_\mu=
(1,0,0,0)$; in SCET, one introduces in addition a light-like
reference vector, $n_\mu = (1,0,0,1)$, in the direction of the jet of
energetic, outgoing, light particles and decomposes all momenta
as\\
\begin{equation}
p_\mu =  n\cdot p\, \frac{\bar
  n_\mu}{2}+\bar{n}\cdot p\, \frac{n_\mu}{2}+p_\mu^\perp\, = p_+^\mu+p_-^\mu+p_\perp^\mu,
\end{equation}
where ${\bar n}_\mu =2v_\mu-n_\mu= (1,0,0,-1)$. The leading order of the
heavy-quark expansion of the diagrams in Figure \ref{fig:scetI} was
first analyzed in \cite{Korchemsky:1994jb}, where it was shown that three
momentum regions lead to singularities in these diagrams. The scaling
of the momentum components $( n\cdot p, \bar{n}\cdot p, p_\mu^\perp)$
in these regions is
\begin{equation}\label{scaling}
\begin{aligned}
\text{hard:}\;\; &(m_b,&m_b,&&m_b), \\
\text{hard-collinear:}\;\; & (\Lambda,&m_b,&&\sqrt{\Lambda\, m_b}), \\
\text{soft:}\;\; &  (\Lambda,&\Lambda,&& \Lambda).
\end{aligned}
\end{equation}
Interestingly, an intermediate scale $\mu_{hc}=\sqrt{\Lambda\, m_b}$ appears,
whose virtuality lies between the small virtuality $\Lambda\sim
M_B-m_b$ of the external lines and the mass of the heavy quark. The
presence of three scales also manifests itself in the factorization
formula for the inclusive decay rate \cite{Korchemsky:1994jb}
\begin{equation}\label{eq:fact}
d\Gamma \sim H\, J \otimes S \,.
\end{equation}
The function $H$ contains the hard corrections, the jet-function $J$
the contribution from the hard-collinear region and the shape function
$S$ the soft contribution. The symbol ``$\otimes$'' indicates the
convolution of the jet-function with the shape function.  Let me note
that while it is simple to see that the above three momentum regions
are necessary to obtain the expansion of the diagrams in Figure
\ref{fig:scetI}, so far no formal proof has been given that these three
regions are sufficient to obtain the expansion of an arbitrary
multi-loop diagram.

To construct the effective theory, one introduces the soft quark field
$q$, the soft gluon field $A_s^\mu$, the heavy-quark field $h$ and the
hard-collinear quark and gluon fields $\xi$ and $A_{hc}^\mu$.  The
momentum components of these fields scale as in (\ref{scaling}) and
the scaling of the field components can be inferred by inspecting the
propagators of the fields. One constructs the Lagrangian for these
fields in such a way that its Feynman rules give the QCD diagrams
expanded in the soft and collinear momentum regions. The technical
details of the construction of the effective Lagrangian are beyond the
scope of this talk, but let me illustrate some of its features using
the leading order Lagrangian
\begin{multline}
{\cal L}_{\rm SCET}^{(0)}=
\bar \xi \frac{\bar n\!\!\!/}{2}\left(i n \cdot D_{hc}+ g\,  n\cdot
  A_s(x_-)+ i {D\!\!\!\!\!/}_{\perp hc}\frac{1}{i \bar
    n\cdot D_{hc}}\, i {D\!\!\!\!\!/}_{\perp hc} \right)\xi \\
+\bar q\,i D_s\!\!\!\!\!\!\!\!/\,\,\,\,q+
\bar h i v\cdot D_s h+{\cal L}^{(0)}_{\rm YM}\label{eq:SCETLagrangian}.
\end{multline}
The purely soft terms on the second line are exactly the leading order
HQET Lagrangian. The hard-collinear quark Lagrangian has a more
complicated form because not all four components of
a hard-collinear quark field scale with the same power of
$\Lambda/m_b$. One can split the original Dirac field into
two-component fields $\xi$ and $\eta$
\begin{equation}
\psi_{hc}=\left(\frac{n\!\!\!\!/\, \bar n\!\!\!\!/}{4} + \frac{\bar n\!\!\!\!/
\,  n\!\!\!\!/}{4}\right)\psi_{hc}= \xi+ \eta
\end{equation}
so that $n\!\!\!\!/\xi=\bar n\!\!\!\!/\eta=0$.  The field $\eta$ is
suppressed by a power of $(\Lambda/m_b)^{1/2}$ with respect to $\xi$
and can be integrated out. Since no approximation is made when $\eta$
is integrated out, the hard-collinear part of the Lagrangian
(\ref{eq:SCETLagrangian}) is, despite appearances, equivalent to the
usual QCD quark Lagrangian. Finally, let me discuss the
interaction term $g\,\bar \xi(x)\frac{\bar n\!\!\!\!\!/}{2}\, {n\cdot
A_s(x_-)}\xi(x)$. The interaction is simpler than the usual QCD
quark-gluon coupling for two reasons. First of all, it only involves a
single component of the soft gluon field.  This is a consequence of
the projection properties of the hard-collinear quark spinor $\bar
\xi\gamma_\perp^\mu \xi=\bar \xi n\!\!\!\!/\, \xi=0$. Secondly, the
soft gluon field only depends on the single coordinate $x_-=\bar n\cdot x
\frac{n_\mu}{2}$. To obtain diagrams expanded in momentum space, one
has to perform a derivative expansion of the effective Lagrangian in
position space \cite{Beneke:2002ph}: in interactions with collinear
fields the soft fields are expanded as
\begin{equation}\label{eq:multipole}
\phi_s(x)=\phi_s(x_-)+x_\perp\cdot\partial \phi_s(x_-)+
x_+\cdot\partial \phi_s(x_-)+\dots\,.
\end{equation}
Alternatively, one can use a hybrid position and momentum space
representation of the collinear fields, the label formalism
\cite{Bauer:2000yr}.  Because of the expansion (\ref{eq:multipole}),
the interaction terms in the Lagrangian (\ref{eq:SCETLagrangian}) are
not translation invariant.  However, the invariance is restored order
by order after including higher order terms. Let me briefly comment on
the gauge transformation properties of the above fields. One can
perform separate gauge transformations on the two gluon fields $A_s$
and $A_{hc}$. The form of these transformations is restricted by the
requirement that they should respect the power counting of the fields
and the fact that the sum of the two gluon fields $A_s+A_{hc}$ has to
transform as the usual QCD gluon field. The explicit form of the gauge
transformations as well as the first three orders of the SCET
Lagrangian can be found in \cite{Beneke:2002ni}. In the label
formalism, the sub-leading Lagrangian was given in
\cite{Pirjol:2002km}.

In order to analyze the inclusive semi-leptonic decay one needs the
weak current $J_\mu$ in the effective theory. At leading order this
current is represented in the effective theory as
\begin{equation}\label{eq:curr}
J^\mu(x) = \sum_{i=1}^3 \int\! ds\,\,C_i(s)\,e^{-i m_b v\cdot x}\,J^\mu_i(x,s)\,.
\end{equation}
The three leading order effective theory current operators are
\begin{equation}\label{eq:current}
J^\mu_i(x,s)=\bar\xi(x+s \bar n) W_{hc}\,(1+\gamma_5)\,
\Gamma^\mu_i\,h(x_-)\,\text{ with } \Gamma^\mu_{1,2,3}=\{\gamma^\mu,v^\mu,n^\mu\}\,.
\end{equation}
The SCET current operators are nonlocal along the light ray in the
$\bar n$-direction. This is a consequence of the fact that the $\bar
n\cdot \partial$-derivative on a hard-collinear field counts as a
quantity of order one and arbitrary powers of this derivative can
appear at any given order in the expansion in $\Lambda/m_b$. The
light-like, hard-collinear Wilson line $W_{hc}$ ensures gauge invariance of the
operator and the phase factor $e^{-i m_b v\cdot x}$ relates the heavy
quark field $h$ to the QCD $b$-quark field. 

To obtain the SCET representation of the current correlator
$T^{\mu\nu}$ one inserts the above representation (\ref{eq:curr}) of the
current into (\ref{eq:Tmunu}). Since the initial and final state
$B$-meson in (\ref{eq:Tmunu}) do not contain collinear partons, one
can perform a second matching step and integrate out the
hard-collinear fields in the products of effective theory currents:
\begin{multline}
J_i^{\dagger\mu}(0,t)J_j^\nu(x,s)=\bar h(0)\, \Gamma_i^\mu\,W^{\dagger}_{hc}\,\xi(t \bar n)\,\bar\xi(x+s \bar n) W_{hc}\,\Gamma^\nu_j\,h(x_-)\\
\rightarrow{\cal J}(x+(s-t){\bar n})\, \bar h(0)\, \Gamma_i^\mu\,\frac{\bar n\!\!\!\!/}{2}\, \Gamma_j^\nu \,h(x_-) +\dots\,.
\end{multline}
The Wilson coefficient ${\cal J}$ that appears in this second step is
called the jet-function. After realizing that the matrix elements of
the operators $\bar h(0)\, \Gamma_i^\mu \frac{\bar n\!\!\!\!/}{2} \Gamma_j^\nu \,h(x_-)$ can be
related to the matrix element of the single operator $\bar
h(0)\,h(x_-)$ by heavy-quark symmetry and rewriting these above
results in momentum space, one obtains the factorization theorem
(\ref{eq:fact}). 

 While the factorization theorem was originally derived
diagrammatically, the SCET analysis has led to several improvements in
our understanding of these decays. First of all, solving the
renormalization group equations in the effective theory at the
next-to-leading order, the perturbative single and double logarithms
of the different scale ratios in the hard and the jet-functions have
been resummed at next-to-leading order and predictions for
the various decay spectra have been obtained \cite{Bauer:2003pi,
  Bosch:2004th}, which are free of the spurious Landau pole
singularities present in earlier work \cite{Korchemsky:1994jb,
  Akhoury:1995fp, Leibovich:1999xf}. Also, because of the
renormalization properties of the shape function, the relation between
its moments and higher order HQET parameters is more complicated than
previously assumed \cite{Bauer:2003pi, Bosch:2004th}.

The SCET formalism has also been used to investigate the partial
$B\rightarrow X_s\gamma$ decay rate with a cut $E_\gamma > E_0$ on the
photon energy .  For low values of this cut, the rate can be
calculated using the standard OPE, while for larger values higher
order effects become large and have to be resummed into the shape
function. The transition between the two regimes is smooth: in the
presence of a photon energy cut, the OPE becomes an expansion in
$1/\Delta$, where $\Delta=m_b-2E_0$ is the energy window available
after the cut. Another effect of the cut is that the rate calculated
in the OPE gets perturbative corrections of order $\alpha_s(\Delta)$ .
For realistic values of the cut the scale $\Delta$ is not much larger
than 1GeV. After an analysis in renormalization group improved
perturbation theory, Neubert finds that, even for the low value
$E_0=1.8\, {\rm GeV}$, higher order perturbative corrections to the
cut rate are sizable, of order 15\% \cite{Neubert:2004dd}. This is
larger than previously estimated and larger than the power corrections
of order 10\%.

As the above discussion shows, the effective theory analysis of the
inclusive decay rate reduces to the analysis of the weak
heavy-to-light current.  This analysis has been extended to
sub-leading order in $\Lambda/m_b$
\cite{Bauer:2001mh,Bauer:2002yu,Lee:2004ja,Bosch:2004cb}. While a
single non-perturbative shape function is sufficient to obtain the
leading order rate, several shape functions appear at the next order.
At leading order, one can determine $|V_{ub}|$ by taking a ratio of
weighted $B\rightarrow X_u \ell \nu$ and $B\rightarrow X_s \gamma$
decay spectra for which the shape function effects drop out.  An
understanding of the corrections to this relation from sub-leading
shape functions is important for a precise determination of
$|V_{ub}|$.

\subsection{Exclusive heavy-to-light decays: SCET$_{\rm II}$}

It turns out that exclusive $B$-decays have a more complicated
structure than the inclusive decays studied in the previous section.
To analyze the decays diagrammatically, one again replaces the mesons
with interpolating currents. For example, in order to analyze the
$B\rightarrow \pi$ form factor of a current $J_\Gamma^\mu(0)$, one uses
the relation
\begin{multline}
  \int\! d^4x\,\int\! d^4y\, e^{i p_\pi x-i p_B y} \langle 0\,|\,T\left\{
    J^\dagger_B(x) J_\Gamma^\mu(0) J_\pi(y)\right\}\, |\,0 \rangle \\=
  \frac{i f_B^{(J)}}{p_B^2-M_B^2} \frac{i
    f_\pi^{(J)}}{p_\pi^2-M_\pi^2}\,\langle \pi(p_\pi)|\,J_\Gamma(0)\,| B(p_B)
  \rangle+\dots \label{eq:LSZ}
\end{multline}
and analyzes the factorization properties of the correlator on the
left-hand side diagrammatically.\footnote{The amplitude on the right
  hand side is part of the double spectral density of the correlator
  with respect to the variables $p_\pi^2$ and $p_B^2$. Factorization
  of the double spectral density is therefore sufficient to establish
  factorization of the amplitude.}

  Here, the currents $J_B$ and $J_\pi$
have the quantum numbers of the $B$-meson and the pion respectively
and associated decay constants $f_B^{(J)}$ and $f_\pi^{(J)}$.  The ellipsis
denotes terms that do not have a pole at $p_B^2=m_B^2$ with $p_B^0<0$
and $p_\pi^2=m_\pi^2$ with $p_\pi^0>0$. 

\begin{figure}
%\begin{center}
\begin{tabular}{ccc}
\raisebox{0.55cm}{$p_B\,$}
 \raisebox{0.0075\textwidth}{\includegraphics[width=0.27\textwidth]{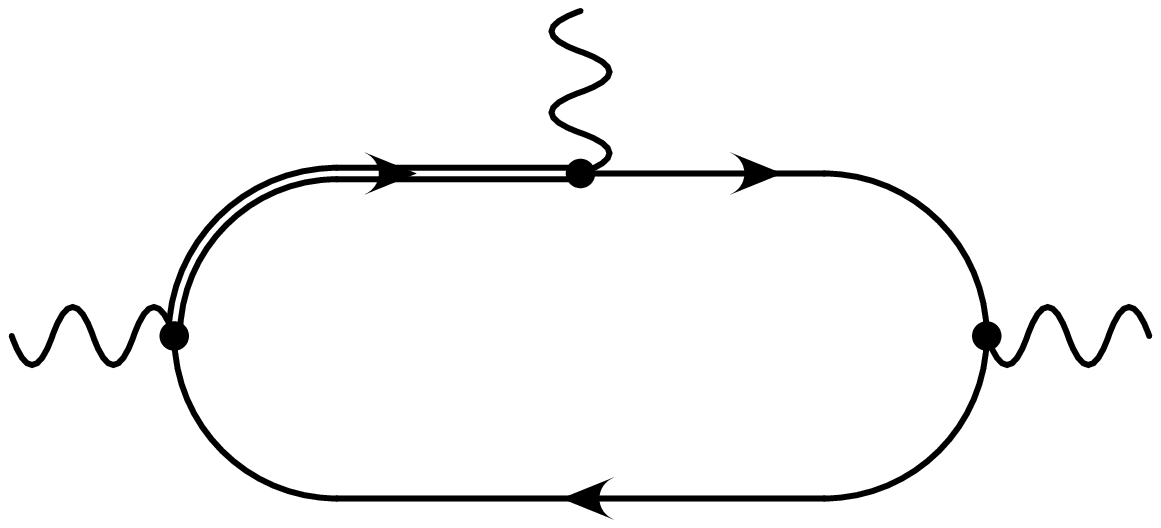}}
\raisebox{0.55cm}{$p_\pi$} 
&\includegraphics[width=0.27\textwidth]{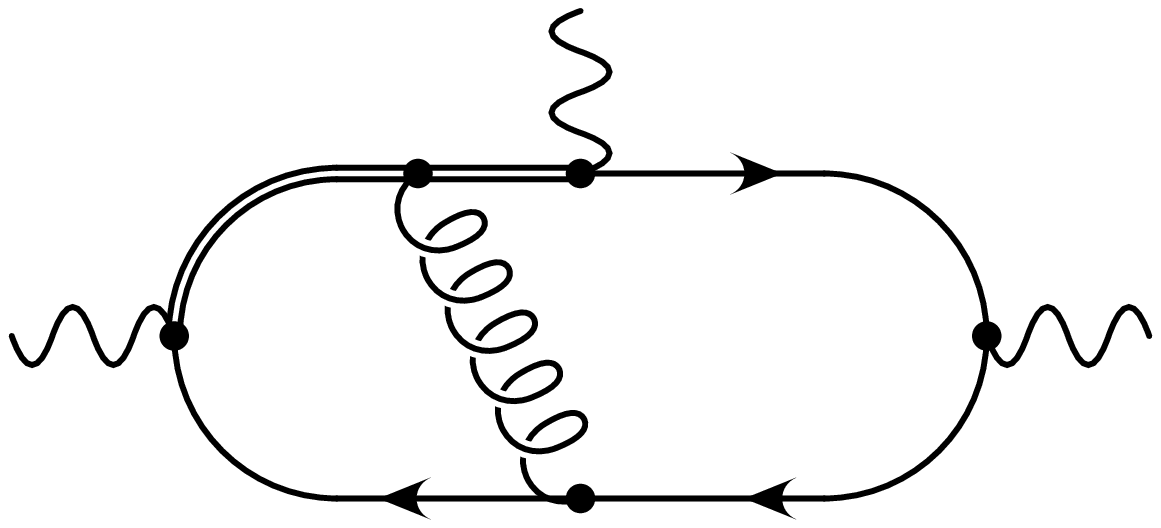}
& \includegraphics[width=0.27\textwidth]{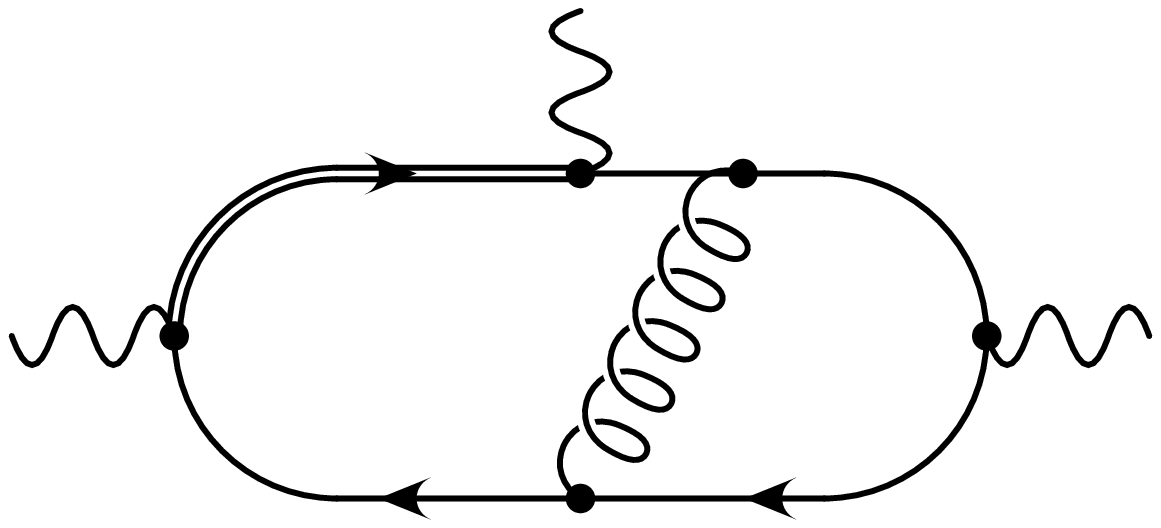}
\end{tabular}
%\end{center}
\caption{Lowest order QCD diagrams contributing to the correlator (\ref{eq:LSZ}). The
 double line denotes the heavy quark.\label{fig:correlator}
}
\end{figure}

The lowest order diagrams contributing to the correlator
($\ref{eq:LSZ}$) are shown in Figure \ref{fig:correlator}. The
following momentum regions are needed to obtain their expansion \cite{Becher:2003qh} 
\begin{equation}\label{scalingII}
\begin{aligned}
\text{hard:}\;\; &(m_b,& m_b,&&m_b), \\
\text{hard-collinear:}\;\; & (\Lambda,&m_b,&&\sqrt{\Lambda\, m_b}), \\
\text{collinear:}\;\; & (\Lambda^2/m_b,&m_b,&&\Lambda), \\
\text{soft:}\;\; &  (\Lambda,&\Lambda,&& \Lambda),\\
\text{soft-collinear:}\;\; &  (\Lambda^2/m_b,&\Lambda,&& \sqrt{\Lambda^3/m_b}),
\end{aligned}
\end{equation}
where the three components denote the scaling of $(n\cdot p,\bar n
\cdot p, p_\perp)$ and $\Lambda$ stands for a hadronic scale in the
problem that is independent of $m_b$.  The occurrence of a collinear
region in addition to the hard-collinear region is easily understood:
the scaling of its components is the same as that of the external pion
momentum. Since $p_\pi^2$ is independent of the $b$-quark mass, the
hard-collinear scaling would not be appropriate. What is surprising
is the presence of an additional region with very low virtuality: the
soft-collinear region with $p^2_{sc}=\Lambda^3/m_b$.  In the effective
theory, these soft-collinear fields represent a low energy interaction
between the collinear and soft sectors. The scaling of their momentum
components is such that they can be emitted and absorbed by both soft
and collinear particles without taking either of them off-shell. An
example of a diagram where a soft-collinear contribution appears is
the first diagram in Figure \ref{fig:correlator}, which gets a
contribution when the spectator quark is soft-collinear and almost all
external momentum flows through the current quarks.

If these soft-collinear fields contribute to a given quantity in
perturbation theory, factorization is explicitly violated. Since they
are the only low energy interaction between the collinear and soft
sectors, the opposite is also true: if they do not contribute to a
given quantity, the quantity factorizes. Clearly, the physics
associated with the soft-collinear fields cannot be calculated reliably in
perturbation theory: as soon as there is such a contribution, the
corresponding quantity will be deemed non-perturbative. The assumption
on which the factorization framework relies is that if one does not find
a violation of factorization in perturbation theory the quantity
indeed factorizes.

Unfortunately, even the seemingly simplest decays, namely the
heavy-to-light meson form-factors, do not factorize in the
heavy-quark limit. At large recoil energies $E\sim M_B/2$, they take
the form \cite{Beneke:2000wa}
\begin{equation}\label{eq:soft_plus_hard}
  F_i^{B\to M}(E) = C_i(E)\,\zeta_M(E) +
  \int_0^\infty\!{d\omega\over\omega}\,\phi_B(\omega)
  \int_0^1\!du\,f_M\,\phi_M(u)\,T_i(E,\omega,u) \,,
\end{equation}
up to corrections suppressed by $1/m_b$. The non-factorizable piece is
given by the non-perturbative function $\zeta_M(E)$ that depends on
the momentum transfer. The remainder factorizes into a convolution of
a hard scattering kernel $T_i(E,\omega,u)$ with the light-cone
distribution amplitudes $\phi_B(\omega)$ and $\phi_M(u)$ of the two
mesons. The coefficients $C_i(E)$ and the kernels $T_i(E,\omega,u)$
are Wilson coefficients of effective theory operators. While the form
factors do not factorize, one obtains relations between different form
factors because the function $\zeta_M(E)$ is independent of the Dirac
structure of the current under consideration \cite{Charles:1998dr}. In
this respect the situation is rather similar to the case of the
$B\rightarrow D$ transition form factors, which in the heavy-quark
limit can be expressed in terms of a single function, the Isgur-Wise
function, up to perturbative corrections \cite{Isgur:1989ed}.

\begin{figure}
  \includegraphics[height=.23\textheight]{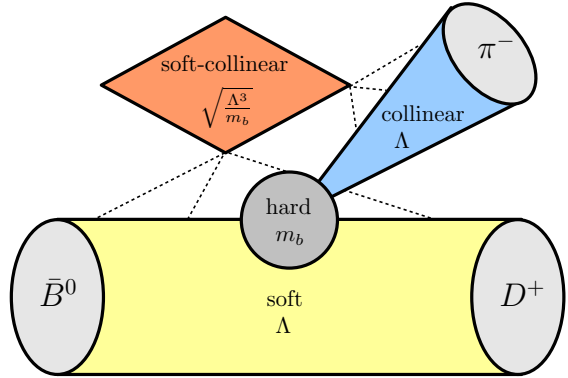}\\
\caption{Momentum regions in the perturbative analysis of the exclusive decay $B\rightarrow D\pi$.\label{fig:SCETII}}
\end{figure}

The heavy-to-light form factors are in a sense sub-leading quantities,
since the superficially leading term in a diagrammatic analysis
happens to vanish. By naive power counting, one would expect the third
diagram in Figure \ref{fig:correlator} to be the leading contribution,
but upon closer inspection one finds that the expansion of all three
diagrams starts at the same order. In the effective theory this
manifests itself in the fact that one needs to include sub-leading
current operators to analyze the form factor to leading order. To
perform the analysis, one first matches QCD onto SCET$_{\rm I}$. The
leading order effective theory currents have been discussed in the
previous section, see (\ref{eq:current}). The reason for the above
mentioned symmetry relations between the different form factors is
that both the heavy and the hard-collinear light quark field are
represented by two component fields. Two types of current operators
appear at next-to-leading order: operators with an additional
derivative $\partial^\mu_\perp$ on the hard-collinear quark and
operators with an additional hard-collinear gluon field
$A_{hc\perp}^\mu$. The operators with an additional derivative have
the same Wilson coefficients as the leading operators and their matrix
elements fulfill the same symmetry relations. The matrix elements of
the remaining sub-leading operators which contain an additional
hard-collinear gluon field $A_{hc\perp}^\mu$ violate the symmetry
relations. To prove the factorization theorem
(\ref{eq:soft_plus_hard}), one matches them onto SCET$_{\rm II}$ and
shows that their contribution to the form factors factorizes.  This
analysis has been carried out and the formula
(\ref{eq:soft_plus_hard}) was established recently
\cite{Bauer:2002aj,Beneke:2003pa,Lange:2003pk}.  Also, the matching
calculations have been performed to one-loop order.  The coefficients
$C_i$ were determined in \cite{Bauer:2000yr, Beneke:2004rc}, and the
two matching steps for $T_i$ were performed in \cite{Hill:2004if,
  Beneke:2004rc, Becher:2004kk}.  The hard scattering kernels contain
single and double (or Sudakov) logarithms of large scale ratios. These
logarithms can be resummed by solving the renormalization group
equations in the effective theory, which was done in
\cite{Hill:2004if}. The analysis in this paper also allows one to
categorize the relations between the different form factors:
generically, one can expect corrections of the order of $\alpha_s$ at
the hard-collinear scale $\mu_{hc}^2 \sim \Lambda\, m_b$. In some
cases, however, the corrections only involve $\alpha_s$ at the hard
scale and some relations do not receive radiative corrections at all
\cite{richardQCD04}. The heavy-to-light form factors are an important
element of the factorization theorem for the phenomenologically
important decays to two light mesons $B\rightarrow M_1\, M_2$. Using
the factorization theorem for these decays \cite{Beneke:1999br},
predictions for the complete set of the almost one hundred $B^-$,
$B^0$ and $B_s$ decay modes are given in \cite{Beneke:2003zv}. These
decays have been investigated using SCET
\cite{Chay:2003zp,Bauer:2004tj}, but a complete analysis is not yet
available.

The decays $B\rightarrow D M$, depicted in Figure \ref{fig:SCETII},
have a simpler structure. In this case an analysis using the effective
theory at leading order is sufficient and the soft-collinear mode does
not contribute in the heavy-quark limit. The amplitude factorizes into
a soft part (the $B\rightarrow D$ form-factor at zero momentum
transfer) and a collinear part which is given by the pion light-cone
distribution amplitude and is convoluted with a hard scattering kernel
\cite{Dugan:1990de,Politzer:1991au,Beneke:2000ry,Bauer:2001cu}. Using
the effective theory framework, also the power suppressed neutral
decay modes $B^0\rightarrow D^{(*)0} M^0$ have been analyzed recently
and it was found that the branching ratios as well as the strong
phases are equal for the decay to the vector and pseudoscalar
$D$-meson \cite{Mantry:2003uz,Mantry:2004pg,blechman}. Analogous
relations also hold for non-leptonic $b\rightarrow c$ decays of
baryons \cite{Leibovich:2003tw}.

%Also,
%factorization for $B\rightarrow \gamma \ell \nu$ has been established
%in 

\section{Summary}

The techniques used to analyze hard-scattering factorization at large
momentum transfer can also be used to analyze decays of heavy to light
hadrons in the heavy-quark limit. While based on the same theoretical
concepts, the factorization theorems for exclusive heavy-to-light
decays are complicated in that the naively leading order term
vanishes and one has to perform the diagrammatic analysis to
sub-leading order. SCET allows one to perform this analysis in a
controlled and transparent way. The effective theory fields correspond
to the momentum regions in which the diagrams develop singularities
and the hard scattering part of these processes is encoded in the
Wilson coefficients of the effective theory operators.  The
resummation of large perturbative logarithms is achieved by solving
the RG evolution equation of the corresponding operators.

Since the hard scale for these decays is given by the $b$-quark mass,
it is numerically smaller than typical momentum transfers in hard
processes at colliders. An understanding of the size of the power
corrections is therefore important. Using the effective theory
framework, these power corrections have been worked out and estimated
for the inclusive semi-leptonic decays. For the exclusive decays, we
have not yet reached the same level of accuracy because the leading
terms in the amplitudes in general already involve the sub-leading
Lagrangian. An important achievement was the analysis of the
heavy-to-light form factors and the proof of the factorization theorem
for this case. Using the same techniques, a complete analysis of the
factorization properties of the charmless two-body decays is within reach.
Some of the power corrections for these decays have been identified
and estimated, but no complete treatment is available to date. While such an
analysis seems difficult, it is important: Our capability to explore
new flavor physics in $B$-decays can only be as good as our
understanding of the strong interaction physics in their decays.

%%%%%%%%%%%%%%%%%%%%%%%%%%%%%%%%%%%%%%%%%%%%%%%%
%% BACKMATTER
%%%%%%%%%%%%%%%%%%%%%%%%%%%%%%%%%%%%%%%%%%%%%%%%

\begin{theacknowledgments} 
  I thank Richard Hill and Matthias Neubert for useful comments on the
  manuscript. Fermilab is operated by Universities Research
  Association Inc.~under Contract No.~DE-AC02-76CH03000 with the U.S.
  Department of Energy.
%This work was supported by the
%Department of Energy under Grant DE-AC02-76SF00515 and by the National
%Science Foundation under Grant PHY99-07949
\end{theacknowledgments}

%% If the bibliography is produced without BibTeX comment out the

\bibliographystyle{aipproc}   % if natbib is available

\end{document}